\documentstyle[11pt]{article}

\input epsf

\global\arraycolsep=1pt
\begin{document}
\def\o#1#2{{#1\over#2}}
\def\bfone{\relax{\rm 1\kern-.35em 1}}
\def\inbar{\vrule height1.5ex width.4pt depth0pt}
\def\IC{\relax\,\hbox{$\inbar\kern-.3em{\rm C}$}}
\def\ID{\relax{\rm I\kern-.18em D}}
\def\IF{\relax{\rm I\kern-.18em F}}
\def\IH{\relax{\rm I\kern-.18em H}}
\def\II{\relax{\rm I\kern-.17em I}}
\def\I1{\relax{\rm 1\kern-.28em l}}
\def\IN{\relax{\rm I\kern-.18em N}}
\def\IP{\relax{\rm I\kern-.18em P}}
\def\IQ{\relax\,\hbox{$\inbar\kern-.3em{\rm Q}$}}
\def\IZ{\relax\,\hbox{$\inbar\kern-.3em{\rm Z}$}}
\def\IR{\relax{\rm I\kern-.18em R}}
\font\cmss=cmss10 \font\cmsss=cmss10 at 7pt
\def\ZZ{\relax\ifmmode\mathchoice
{\hbox{\cmss Z\kern-.4em Z}}{\hbox{\cmss Z\kern-.4em Z}}
{\lower.9pt\hbox{\cmsss Z\kern-.4em Z}}
{\lower1.2pt\hbox{\cmsss Z\kern-.4em Z}}\else{\cmss Z\kern-.4em
Z}\fi}

\hfill    HUTP-95/A026


\hfill   July, 1995

\begin{center}
\vspace{10pt}
{\large \bf
ON FIELD THEORY QUANTIZATION AROUND INSTANTONS
}
\vspace{10pt}

{\sl Damiano Anselmi}

\vspace{4pt}

{\it Lyman Laboratory, Harvard University, Cambridge MA 02138,
U.S.A.}

\vspace{12pt}

{\bf Abstract}
 \end{center}

\vspace{4pt}
With the perspective of looking for experimentally detectable
physical
applications
of the so-called topological embedding, a procedure recently
proposed by the author for quantizing a field theory around
a non-discrete space of classical minima (instantons, for example),
the physical implications are discussed in a ``theoretical''
framework, the ideas are collected in a simple logical
scheme and the topological version of the Ginzburg-Landau
theory of superconductivity is solved in the intermediate
situation between type I and type II superconductors.
\eject

\section{Introduction and motivation}
\label{intro}
\setcounter{equation}{0}

So far, the nonperturbative
aspects of quantum
field theory have been mainly investigated
in the context of supersymmetry.
This research began with the
computation of certain exact ``topological''
correlation functions that are
independent of the positions of
the local observables (condensates) \cite{rossi}.
In many cases, supersymmetry greatly
reduces the computational effort
and various quantities of a topological
nature can be identified.
In non-supersymmetric theories, for example,
primed determinants do not
simplify between bosons and fermions
and have to be calculated explicitly.
Although technically involved, this is not impossible,
as shown by t'~Hooft in ref.\ \cite{thooft}.
At that time, however, no well-defined
exact amplitude
was identified and the convergence of the
integration over the moduli space
remained an open issue
\cite{thooft,rajaraman,coleman}.

It is natural to ask ourselves, nowadays, whether
it can be fruitful to reconsider non-supersymmetric gauge field
theory and explore the viewpoints that are offered by the new
theoretical devices at our disposal.

In ref.\ \cite{me}, a proposal was made to expand perturbatively a
quantum field theory in the
topologically nontrivial sectors,
or, more generally,
when the minima of the classical action
are not a discrete set,
but a {\sl moduli} space ${\cal M}$.
One needs to separate conveniently
the integration over ${\cal M}$, which
remains a nonperturbative issue,
from the integration
over the quantum fluctuations
``perpendicular'' to ${\cal M}$, that,
instead, can be treated perturbatively.
The topological field theory associated
with ${\cal M}$ is suitably embedded into the physical theory, so
that the
physical theory
is viewed as a perturbative expansion around the topological one.
For example, QCD is formulated
as an expansion around topological Yang-Mills theory.
Nothing changes in the topologically trivial
sector with respect to the usual approach, but many problems
encountered so far in the topologically nontrivial sectors
can be easily bypassed and nonperturbative
quantities can be defined
and computed exactly.
The insertion of suitable topological
observables ${\cal O}_\gamma$
guarantees the convergence of
the integration over ${\cal M}$.

The key idea behind the topological embedding
is very simple and natural.
One can say that the space ${\cal M}$, around which one would
like to define perturbation theory,
enjoys an enhanced gauge symmetry, which is precisely of a
topological nature.
Indeed, in the problem under consideration any
two ${\cal M}$-configurations that are
continuously deformable into each other
have to be considered equivalent.
So, the troubles
found in the past
with the ${\cal M}$-integral
are here viewed as due to an extra gauge symmetry
that was not gauge-fixed.
The ${\cal O}_\gamma$-insertions gauge-fix
the extra symmetry. Effectively, they act as projectors onto some
special point
$m\in{\cal M}$,
the Poincar\`e dual of the differential form $\prod_i{\cal
O}_{\gamma_i}$.
Thus, the topological embedding is a sophysticated way to
reduce perturbation theory around a moduli space
${\cal M}$ to perturbation theory around a point $m\in{\cal M}$,
which
is the quotient of ${\cal M}$ by the topological symmetry.
The topological amplitudes classify the inequivalent
ways of doing this consistently.

The first  nontrivial consequence is that the topological quantities
have a role in the physical theory.
This should have deep implications,
even at the qualitative level,
that hopefully will be compared with experiment
in a non distant future.
The purpose of this paper is precisely
to begin a deeper exploration
of this fascinating subject
in order to look for the physical meaning of the
topological aspects of quantum field theory.

Some physical implications of the topological embedding
can be clarified immediately, in a ``theoretical'' framework.
In ref.\ \cite{me}, the concrete
example of pure non-abelian Yang-Mills
theory was considered.
The first difficulty is to imagine what the topologically nontrivial
sectors of
QCD are in nature, since our physical intuition
is limited, at least at present.
Soon or later, some experiment should reveal them.
For now, it is very convenient to stimulate our imagination
with an analogy. There is another interesting physical theory,
closer to experiments, in which
the nontrivial topological sectors could play a
crucial role: it is the Ginzburg-Landau theory of superconductivity,
that is a $U(1)$ gauge theory
with a Higgs charged scalar.
At a very special value ($\lambda=1$)
of the parameter $\lambda$ that
distinguishes type I ($\lambda<1$) form type II ($\lambda>1$)
superconductors,
the theory admits ``instantons'', whose proper name is in fact
``vortices''. The instanton number is the vorticity.
The various topological sectors are thus labelled by
the number of magnetic flux units penetrating the superconductor,
a number that can be measured:
this is a very clear
picture of the physical meaning of the topologically nontrivial
sectors.
It suggests that any event (amplitude) is naturally placed in a {\sl
specific}
topological sector, and does not receive, as it is commonly believed,
contributions from all the topological sectors: either one flux unit
penetrates the superconductor, or two, or zero \ldots,
but not all contemporarily. The same argument should apply to QCD and
to
any other similar case.
This is a basic feature of the topological embedding. Sometimes, to
avoid
confusion, I shall call QCD$^*$ the theory treated with the new
approach.

Another important implication,
that is stressed
very much in the present paper, is that it is not sufficient,
at the quantum field
theoretical level, to specify the vorticity or the instanton number,
i.e.\ what we can call the {\sl classical} topological sector,
but even when this is specified, there are discrete inequivalent
possibilities, represented by the topological observables
${\cal O}_\gamma$ that are inserted in order to make the ${\cal
M}$-integration
convergent.
We say that such possibilities characterize
different {\sl quantum} topological sectors.
Correspondingly, the topological amplitudes are called the quantum
topological
properties
of the instantons.
Of course, the (classical and quantum) topological sectors represent
metastable
configurations, the transition from one sector to
another one requiring a finite perturbation.

In the example of pure non-abelian Yang-Mills theory
the Pontrjiagin number $k$
is the classical
topological property of the instantons, while the generalized
multilink
intersection
theory uncovered in ref.\ \cite{me} classifies the quantum
topological
properties
of the BPST instanton \cite{bpst}.
In the case of superconductivity, the classical property is the
vortex number,
while the quantum properties are classifed in the second part of the
present
paper,
where the topological version of the theory is solved.

So, to compare the experimental results with the predictions
of the theory it is necessary to specify
the quantum topological sector where
the experiment takes place.
Being a global (because topological) property (i.e.\
it is sensible to the {\sl boundary} of the spacetime manifold),
the quantum topological sector
describes the interaction between the quantum fluctuations
and the experimental apparatus:
it is meaningless, in the topologically nontrivial sectors,
to speak about {\sl free}
asymptotic states and the
fluctuations over the instanton background interact with the
instanton background itself.

Recently, another relation between topological field theory and
quantum field
theory
has been proposed. In ref.\ \cite{martellini} Cattaneo,
Cotta-Ramusino,
Gamba and Martellini
formulated QCD as an expansion around a B-F theory.
The idea is as follows. In the
first order formalism, the action
of pure Yang-Mills theory
is written as
\begin{equation}
\int {\rm Tr}[ B\wedge F_A]-g^2\int {\rm Tr}[ B^2].
\end{equation}
Then, a perturbation around
$g=0$ is indeed a perturbation around a B-F theory.
Notice that this trick does not work after the rescaling
$A\rightarrow gA$.
In other words, this formulation treats instantons on the same
footing as any
configuration
with zero Pontrjiagin number, a feature also shared by the
topological
embedding.

An interesting problem is to understand
the correspondence
between the two formulations of QCD that I have just recalled
\cite{progr}.
Abstractly, one is naturally lead to conjecture
that there exists a deep
relation (that I call {\sl the topological map})
between topological Yang-Mills theory
and the B-F theory (see fig.\ 1).
This relation may seem a bit
unplausible, at first sight, and, indeed, has never been conjectured
so far.
However, we already have a strong evidence in favour of it.
It is well-known that link numbers are natural topological invariants
of the B-F theories. On the other hand,
the results collected in
ref.\ \cite{me} (firstly discovered in \cite{anomali})
show that the amplitudes of topological Yang-Mills theory
with the BPST instanton
are also link invariants. This was quite unexpected, indeed.
Presumably, the topological map
passes very nontrivially through the physical
theory (i.e.\ QCD itself), so that understanding abstractly
the topological map could give insight to uncover other
nonperturbative
aspects of QCD \cite{progr}.

\let\picnaturalsize=N
\def\picsize{4.0in}
\ifx\nopictures Y\else{\ifx\epsfloaded Y\else\fi
\global\let\epsfloaded=Y
\centerline{\ifx\picnaturalsize N\epsfxsize \picsize\fi
\epsfbox{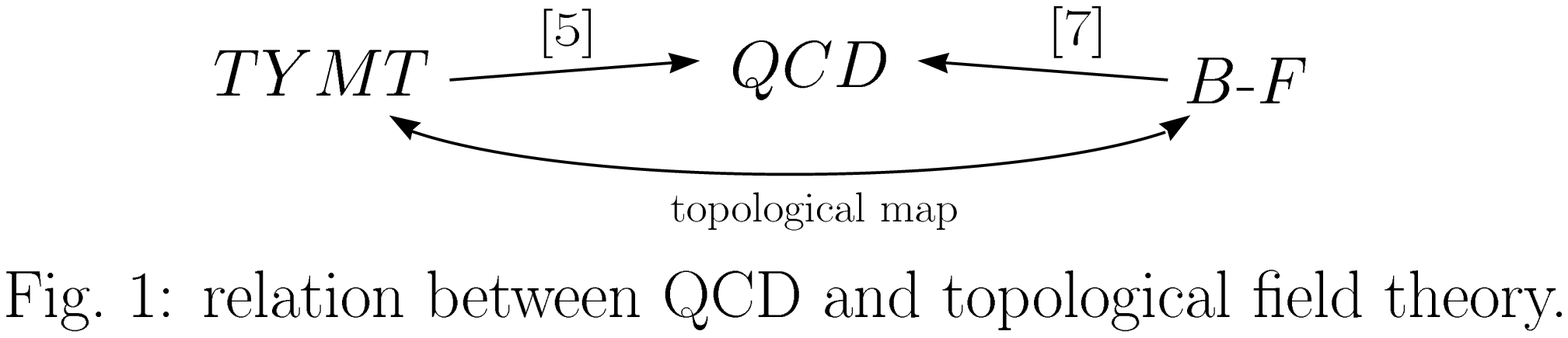}}}\fi

As we see, there are many interesting open problems,
both of a mathematical and a physical nature.
On the physical side, surely the more attractive one,
other questions are:

i) what are the quantum topological sectors of ref.\ \cite{me} in
nature?

ii) how to detect the non-abelian analogue of the Aharonov-Bohm
effect
described in ref.\
\cite{me}?

iii) can we find a second quantization or a statistical distribution
collecting the quantum topological sectors
into a general picture\footnotemark\footnotetext{This
interesting question was raised to me by P.\ Fr\`e.}?

iv) can topological field theory describe properties of zero
temperature
phenomena?

v) can interactions with the boundary be adequately described by the
topological observables and the topological embedding?

vi) is the quantum Hall effect related to all of this?

I shall not answer all these questions in the present paper, but many
interesting remarks will
be made along with the discussion. Moreover,
it is very useful to describe
the various aspects of the new
approach from
different viewpoints. The first part of
the paper is devoted to this.
The long-range aim of the investigation is to
compare the consequences
of the topological embedding with experiment. Therefore,
it is necessary to identify the most convenient theories, situations
and physical quantities. It would not be upsetting if topological
field
theories
turned out to give good descriptions of phenomena
where the quantum fluctuations are naturally frozen,
like at zero temperature.
Having this in mind, I consider, in the
second part of the paper, the Ginzburg-Landau theory
of superconductivity and solve its topological version in the
intermediate case between type I and type II superconductors.
This investigation could stimulate the experimental study of
superconductors in
this very special
situation.

Point vi) is suggested by the observation that
the quantum Hall effect
is a zero temperature phenomenon characterized by a quantity,
the filling factor, that takes integer and fractional values. Such
values have
all the features of the topological quantities and could be a direct
or
indirect trace of the quantum topological sectors.

As far as point iii) is concerned,
one might ask
whether there is something missing in the
theory, and why, otherwise, the topological
embedding is unable to describe transitions
among different quantum topological sectors.
To clarify this point, I stress that
the topological embedding is not an
exact treatment of the theory, but a way
of defining perturbation theory around
non-discrete spaces of minima. As such,
the topological embedding
is perfectly satisfactory to describe
phenomena involving the fluctuations
over the (classical and quantum) background, but it fails
to describe nonperturbative
transitions. The complete theory,
nevertheless, is expected to contain the answer
to this. Yet, it could be very interesting
to try to supplement the topological embedding
with something
like a second quantization or a generalized statistical distribution,
in order to get closer to the
complete theory.

The paper is organized as follows. In section \ref{zero} the basic
facts about
the topological embedding are recalled, the
experimental normalization of coupling constants
is analysed in the new approach and the concrete meaning of the
notion of quantum topological sector  ${\cal Q}$ is developped.
In section \ref{qtp} the quantum topological properties of
instantons computed so far are summarized.
In section \ref{uno} the topological version of the
Ginzburg-Landau theory of superconductivity is solved.

\section{Quantum field theory around a non-discrete space of minima}
\label{zero}
\setcounter{equation}{0}

In this section, the proposal of ref.\ \cite{me}
for quantizing a theory when the action has not
a unique minimum, but a
non-zero dimensional space ${\cal M}$ of minima, is briefly recalled.
In subsection \ref{norm} the normalization of physical
constants in the quantum topological sectors is elucidated.
This gives a concrete interpretation of the physical meaning of the
${\cal O}_{\gamma}$-insertions (see question i) of the introduction).

Since there is no way to privilege {\sl one} solution
to the field equations, the classical theory is not satisfactory.
In quantum field theory, on the other hand,
the functional integral should bypass the problem. However, in
general,
the ${\cal M}$-integral is not guaranteed to converge
\cite{thooft,rajaraman,coleman}.
The question that one has to answer in order to correctly define
the quantum theory is: {\sl how} to do the integral over
${\cal M}$? How many inequivalent possibilities are there? What do
they mean
physically?

The {\sl topological embedding} proposed in ref.\ \cite{me} in order
to
overcome the problem correctly is a way of expanding the true
theory around the topological version of the same theory.
Notice that there is a topological version of {\sl any} quantum field
theory.
Moreover, the topological theory can be defined for {\sl any} given
finite
dimensional subspace ${\cal M}$ of the total functional space.
In this respect, there is nothing special with the objects that we
usually call
``instantons''.
Topological field theory is a useful device to compute the quantum
topological
properties of any given ${\cal M}$. It is the physical theory that
selects the
interesting ${\cal M}$'s (the minima of the classical action).
Keeping this in mind, I shall nevertheless refer to  ${\cal M}$
as to the space of  ``instantons''.

Here are the basic features of the topological embedding.
To be concrete, I take Yang-Mills theory.
The gauge field $A$ is written as $A=A_0+gA_q$, $A_0$ parametrizing
the space
${\cal M}$ and $A_q$ denoting the quantum fluctuation. In the
topologically
trivial sector
$A_0=0$, so that $A=gA_q$  corresponds to the replacement
$A\rightarrow gA$
that defines the usual perturbative approach.
The Yang-Mills
BRST algebra for $A$ is written as the semidirect product of the
topological
BRST algebra for $A_0$ and the consequent remnant for $A_q$. This
formulation,
that otherwise would be nothing else but a refined version of
the Faddeev-Popov procedure of ``introducing 1''
\cite{faddeev,patrasciou},
has the advantage of providing a set
of topological observables ${\cal O}_\gamma$
(constructed with $A_0$ and
promoted now to physical observables of the complete theory),
that can be used to make the integration over the moduli space
well-defined.
The physical amplitudes are thus defined
as
\begin{equation}
\ll A_q(x_1)\cdots A_q(x_n)\gg_{\cal Q}\equiv
<\prod_i{\cal O}_{\gamma_i}\cdot A_q(x_1)\cdots A_q(x_n)>.
\label{21}
\end{equation}
${\cal Q}$ is a label identifying completely the topological sector
where
the experiment takes place. One is lead to define a notion
of {\sl classical} topological sector and a notion of {\sl quantum}
topological
sector, as follows.

i) The classical topological sector is identified by the value
of classical topological invariant, in our case the
instanton number $k$, contributing to the amplitude.
For (\ref{21}) it is the total ghost number of the ${\cal
O}_\gamma$-insertions:
\begin{equation}
|k|=\sum_i{\rm gh}\#\,[{\cal O}_{\gamma_i}].
\end{equation}

ii) The quantum topological sector ${\cal Q}$ is identified
by the specific set of ${\cal O}_\gamma$'s that have been inserted.

The classical topological invariant is a common property of {\sl any}
single
instanton
in the moduli space ${\cal M}$, not a property of the {\sl space}
${\cal M}$ of instantons. The amplitudes of the topological theory,
on the
other
hand, arise as integrals over ${\cal M}$.
It is proper of a {\sl quantum} theory to deal,
via the functional integral, with the space of configurations
and not with single configurations. This is the reason why the
instanton number
is here called the {\sl classical} topological invariant, while the
amplitudes
of the
topological field theory are called
the {\sl quantum} topological invariants of the instanton.

When $n=0$ in (\ref{21}) (i.e.\ when no functional derivative is
taken with
respect to the source $J_q$ associated to the quantum fluctuations
$A_q$
and $J_q$ is set to zero),  the amplitude
$\ll 1\gg_{\cal Q}$ is proportional to the pure
quantum topological invariant $<\prod_i{\cal O}_{\gamma_i}>$
(see subsection \ref{norm} for the detailed justification of this).
It plays the role, in the topologically nontrivial sectors,
that is played by the partition function
in the topologically trivial sector
(which is simply equal to 1, at $J_q=0$, by normalization).
As such, $\ll 1\gg_{\cal Q}$ is not detectable
in a direct way. Indeed, a ``particle'' is an excitation {\sl above}
a background, not a property of the background.
What one can concretely do, instead, is to study gluon scattering
{\sl over}
the given
quantum background ${\cal Q}$ and compare the predictions with the
results found in the topologically trivial sector. The effect of the
quantum background should be enough for an eventual
(perhaps only hypothetical, for now) experimental test.

\let\picnaturalsize=N
\def\picsize{3.0in}
\ifx\nopictures Y\else{\ifx\epsfloaded Y\else\fi
\global\let\epsfloaded=Y
\centerline{\ifx\picnaturalsize N\epsfxsize \picsize\fi
\epsfbox{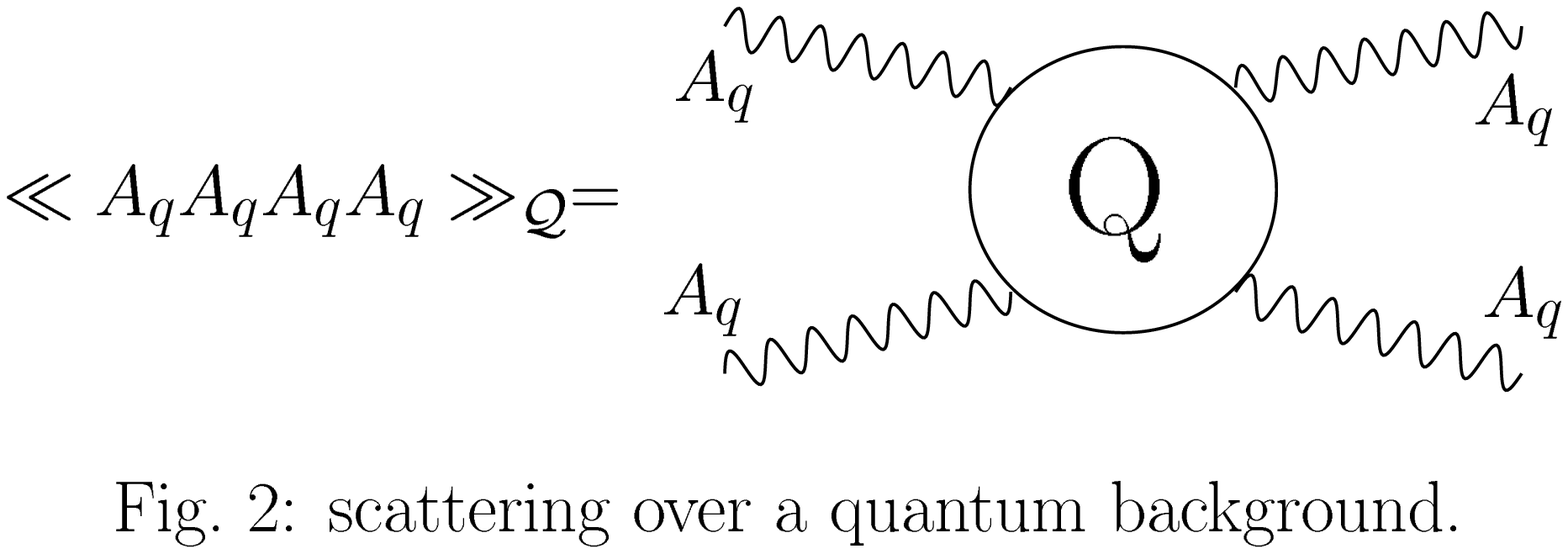}}}\fi

I stress that the topological embedding
(and the approach of ref.\ \cite{martellini},
on the side of  B-F theories) is the answer to the question about the
physical
relevance
of topological field theory, a subject that, surprisingly, had not
been
considered seriously
in the previous literature.

\subsection{Normalization of physical constants in QCD$^*$}
\label{norm}

Let us write the generic amplitude of the ${\cal Q}$-quantum
topological sector
of QCD$^*$ in the form
\begin{equation}
G^{(n)}_{\cal Q}(p,q,g_{\cal Q})=\int_{\cal M}d\rho\,{\cal
Q}(\rho,q)\,\gamma_{(n)}
(p,\rho,\Lambda,g_R,Z,Z_g).
\end{equation}
Here, most terms are symbolic. $\rho$ often denotes the entire set
of moduli, but it is only on the scale that we need to focus on.
${\cal Q}(\rho,q)$ represents the (explicitly known
\cite{anomali,me})
insertion of topological observables $\prod_i{\cal O}_{\gamma_i}$,
that depend on the moduli and on the $\gamma_i$. The
$\gamma_i$-dependence is here symbolically denoted by the momentum
$q$,
because we are working in momentum space.
In general, this dependence
affects any dynamical amplitude.
Only the topological amplitudes are $q$-independent.
$\gamma_{(n)}$ is the perturbative part of the amplitude, regularized
with a cut-off
$\Lambda$. $p$ are the external
$A_q$-momenta and $n$ is the number of external $A_q$-legs.
$g_{\cal Q}$ is the measured coupling constant, at a certain
reference
scale that will be specified. $g_R$, instead, has no direct physical
meaning.
It represents the correct expansion parameter for the perturbative
part
$\gamma_{(n)}$. The relation between $g_{\cal Q}$ and $g_R$ will
come out of the argument.
$Z$ and $Z_g$ are the wave function and coupling renormalization
constants.
They depend on $g_R$, $\Lambda$ and a certain reference scale
$s$.

Now, we have to understand the meaning of $q$, $s$, $g_{\cal Q}$
and $g_R$ and say what is measured and how.
In particular, ${\cal Q}$ fixes the quantum topological sector, but
what is the
meaning of $q$?

I have recalled, in the introduction, that in QCD$^*$
each event takes place
in a single topological sector and does not receive contributions
from
any sector.
The first consequence of this fact is that the physical
normalization of the coupling constant
is not the same in any sector.
The second consequence is that it is meaningless to say that
instanton
contributions are suppressed: in the topologically
nontrivial sectors they are the entire story.

$Z$ and $Z_g$ should be such that $\gamma_{(n)}$ is convergent in
$\Lambda$.
If we do not require this {\sl before} the ${\cal M}$-integration,
we can loose the powerful theorems about the classification of
divergences.
On the other hand, since $g_R$ has no direct physical meaning,
its reference scale $s$ does not need to have a physical meaning.
Moreover, in $\gamma_{(n)}$ there is already a scale
(that is an ``external'' scale, from the point of view of
$\gamma_{(n)}$),
namely $\rho$. Consequently, the natural choice is $s=\rho$.

After the $\Lambda\rightarrow \infty$ limit, we have
\begin{equation}
G^{(n)}_{\cal Q}(p,q,g_{\cal Q})=\int_{\cal M}d\rho\,{\cal
Q}(\rho,q)\,\gamma_{(n)}
(p,\rho,g_R).
\end{equation}
Now, let us consider  $n=3$, namely the vertex function that fixes
the coupling
constant. It is reasonable
to look at $q$ as the {\sl generalized} reference
scale at which the physical parameters are normalized
and this is consistent with the interpretation according to which
${\cal Q}$ specifies
some kind of interaction with the experimental apparatus.
Thus, for $p=q$ we write
\begin{equation}
G^{(3)}_{\cal Q}(q,q,g_{\cal Q})=\int_{\cal M}d\rho\,{\cal
Q}(\rho,q)\,\gamma_{(3)}
(q,\rho,g_R)\equiv g_{\cal Q}(q).
\end{equation}
$q$ and $g_{\cal Q}(q)$ have physical meaning:
$q$ as a chosen (generalized) scale, $g_{\cal Q}(q)$ as a measured
number at
the scale $q$.
On the other hand, $g_{\cal Q}(q)$ is a function of $g_R$. So,
indirectly we have $g_R=g_R(q)$, which fixes $g_R$. The running
coupling
constant is
\begin{equation}
g_{\cal Q}(p)\equiv G^{(3)}_{\cal Q}(p,q,g_{\cal Q})=\int_{\cal
M}d\rho\,{\cal
Q}(\rho,q)\,\gamma_{(3)}
(p,\rho,g_R).
\end{equation}
Its behaviour could be different from the running of the coupling
constant
in the topologically trivial sector.

Now, let us take $n=0$, so that no external momentum $p$ is present.
Since $\gamma_{(0)}$ is dimensionless, it cannot depend on $\rho$.
This means that
\begin{equation}
\ll 1\gg_{\cal Q}=G^{(0)}(q,g_{\cal Q}(q))=
\gamma_{(0)}(g_{\cal Q}(q))\int_{\cal M}d\rho\, {\cal Q}(\rho,q)=
\gamma_{(0)}(g_{\cal Q}(q))<\prod_i{\cal O}_{\gamma_i}>,
\end{equation}
as desired. In the last term, $<\ldots >$ refers to the pure
topological
theory.
$\ll 1\gg _{\cal Q}$ is the partition function in the ${\cal
Q}$-sector.

The above normalization prescription is not in contraddiction with
the
usual one for the topologically trivial sector. In that case a
fictitious
intermediate scale $\rho$ can be also introduced.
Then, ${\cal Q}(\rho,q)$ is $\delta(\rho-1/q)$: this illustrates the
operation
with which the observer fixes a definite reference scale $q$. The
``topological
invariant'' is simply $\int d\rho \,\delta(\rho-1/q)=1$.

The analysis of this subsection offers a clear interpretation of the
quantum topological sectors ${\cal Q}$:
they classify the ways in
which the observer can fix the reference momenta in order to
normalize the fundamental parameters of the theory, which is
the notion of ``generalized reference scale''.

The simplest example of non-topological amplitude is perhaps the
propagator of a scalar field in the fundamental representation of
$SU(2)$ above
the BPST instanton background. In this case, the
perturbative Green function is (see appendix D of ref.\ \cite{rossi}
for
details)
\begin{equation}
\gamma_{(2)}(x,x^\prime,\rho,x_0)={i\over 4\pi^2}{1\over
(x-x^\prime)^2}{\rho^2+(x-x_0)^\mu(x^\prime-x_0)^\nu e_\mu\bar
e_\nu\over
[\rho^2+(x-x_0)^2]^{1/2}[\rho^2+(x^\prime-x_0)^2]^{1/2}}\,\gamma_{(0)}
(g_{\cal
Q}).
\label{propa}
\end{equation}
As before, the factor $\gamma_{(0)}(g_{\cal Q})$ is due to the
integral over
the gauge field $A_q$, with no $A_q$-external legs. The behaviors of
(\ref{propa}) for $\rho\rightarrow 0, \infty$, and for
$x_0\rightarrow
x,x^\prime,\infty$ show that the convergence of the ${\cal
M}$-integral is
preserved.
A complete amplitude is, for example,
\begin{equation}G^{(2)}_{\cal Q}(x,x^\prime,y,z,g_{\cal Q})=
{1\over  (x-x^\prime)^2}\int_{\cal M}\omega^{(4)}_y \,\omega^{(1)}_z
(x-x^\prime)^2\,
\gamma_{(2)}(x,x^\prime,\rho,x_0),
\end{equation}
where $\omega^{(4)}_y$ is given in formula (4.13) of ref.\
\cite{anomali} and
corresponds to
the local observable, while $\omega^{(1)}_z$ is given in formula
(2.18) of
ref.\ \cite{me}
and is a nonlocal observable integrated over a 3-sphere.

\section{The quantum topological properties of  the instantons}
\label{qtp}
\setcounter{equation}{0}

In this section I briefly recall how link numbers appear in
topological Yang-Mills theory with the BPST instanton
and what happens in the other topological field
theories that have been solved explicitly so far. The purpose of this
section
is to collect the essential features of the matter, while, for the
detailed
proofs and calculations,
the reader should check ref.s \cite{anomali,me}.

The topological observables ${\cal O}_{\gamma_i}$ correspond to
closed
differential forms $\omega_{\gamma_i}$ on the moduli space
${\cal M}$. In the interior of ${\cal M}$ such forms are also exact
and we can define $\Omega_{\gamma_i}$'s such that
$\omega_{\gamma_i}=d\Omega_{\gamma_i}$. We can thus write
\begin{equation}
{\cal A}=<\prod_i{\cal O}_{\gamma_i}>=
\int_{\cal M}\prod_i\omega_i=\int_{\partial{\cal M}}
\Omega_1\prod_{i\neq 1}\omega_i.
\end{equation}
Now, $\omega_{\gamma_i}$ are generated by ${1\over 16\pi^2}
\hat F^a\hat F^a$ \cite{anomali}, which can also be written as $\hat
d \hat C$,
$\hat C$ being the BRST extended Chern-Simons form,
while $\Omega_{\gamma_j}$ are generated by $\hat C$.
On $\partial {\cal M}$, i.e.\ when $\rho\rightarrow 0$ and $d\rho$ is
set to
zero
the explicit solution elaborated in ref.s \cite{anomali,me}
gives, not surprisingly,
\begin{equation}
{1\over 16\pi^2}
\hat F^a\hat F^a(x)\rightarrow{1\over 4!}\delta(x-x_0)\,dV(x-x_0)=
-{1\over  4!\pi^2} \hat d\,
\partial_\mu{1\over (x-x_0)^2} \,d\sigma_{\mu}(x-x_0),
\end{equation}
so that
\begin{equation}
\omega_{\gamma_i}\rightarrow \sim \int_{\gamma_i}\delta(x_i-x_0),
\quad\quad
\Omega_{\gamma_j}\rightarrow \sim \int_{\gamma_j}\partial
{1\over (x_j-x_0)^2}.
\end{equation}
Consequently,
\begin{equation}
<\prod_i{\cal O}_{\gamma_i}>\sim\int_{\IR^4}
dx_0 \int_{\gamma_1}\partial
{1\over (x_1-x_0)^2}\prod_{i\neq 1}\int_{\gamma_i}
\delta(x_i-x_0)\sim
\backslash\!\!\!\slash (\gamma_1,\ldots\gamma_n).
\end{equation}
This facts show that there is a very simple generalization of the
concept of
linking;
it is the so-called multilink invariant
$\backslash\!\!\!\slash (\gamma_1,\ldots\gamma_n)$
that can be reduced to the
usual link number between one
chosen submanifold, say $\gamma_1$, and the intersection among the
other ones,
the result being independent of the chosen $\gamma_1$:
\begin{equation}
\backslash\!\!\!\slash (\gamma_1,\ldots
\gamma_n)=\backslash\!\!\!\slash (\gamma_1,\cap_{i\neq 1} \gamma_i).
\end{equation}
In the formulas sketched above, the symbol $\sim$
means that numerical factors have been neglected.
One can check \cite{anomali,me}, nevertheless, that the
link numbers turn out to be normalized correctly.

Now, due to the topological embedding, the link invariants
$\ll 1\gg_{\cal Q}$ just recalled are also exact QCD amplitudes.
Therefore, since we expect that QCD exhibits confinement at
the nonpreturbative level,
it is natural to ask ourselves whether the above results are
compatible with it or not.
Indeed they are, if we interpret them as
a non-abelian analogue of the Aharonov-Bohm effect \cite{me}.
In the investigation of the physical relevance of the topological
quantities,
the Aharonov-Bohm effect is very important, because it
is a noticeable example of an experimental phenomenon in which a
purely
topological quantity (precisely a link number) plays a central role.
The link number reveals that the magnetic field
is trapped inside the (closed, circular) solenoid $\Gamma$, like in
Fig.\ 3.
$\Gamma$ can be conveniently idealized to a closed circle,
corresponding to a single magnetic force line.
Since $\ll 1\gg_{\cal Q}$ are the only exact nonperturbative QCD
amplitudes
that we possess today, they can be viewed
as a trace that non-abelian Yang-Mills theory confines,
in the sense that they are consistent with this fact and
no exact result available at present is in contraddiction with it.
The same cannot be said of the link numbers in QED, of course:
it is sufficient to open the solenoid to deconfine the field.
In the realm of the amplitudes $\ll 1\gg_{\cal Q}
\propto<\prod_i{\cal
O}_{\gamma_i}> $,
the observables ${\cal O}_\gamma$
are necessarily associated
to {\sl closed} $\IR^4$-submanifolds $\gamma_i$:
opening them would deconfine the field,
but this is forbidden by the gauge invariance.
The intuitive picture that we have just worked out also
suggests that the closed submanifolds $\gamma$
can be viewed as effective color force
{\sl lines, surfaces, 3-spheres,  etc.}

\let\picnaturalsize=N
\def\picsize{4.0in}
\ifx\nopictures Y\else{\ifx\epsfloaded Y\else\fi
\global\let\epsfloaded=Y
\centerline{\ifx\picnaturalsize N\epsfxsize \picsize\fi
\epsfbox{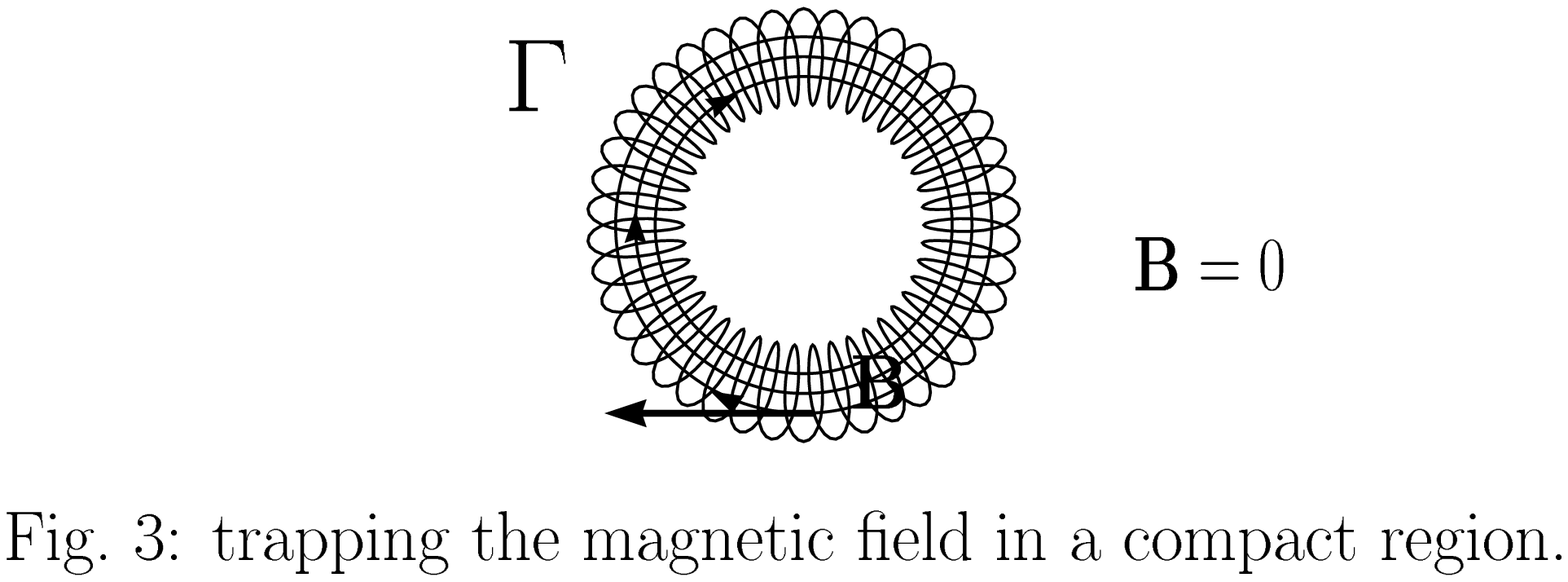}}}\fi

Another important consequence of the
investigation of ref.\ \cite{me} is that an observable ${\cal O}$
can be `sensitive' to the positions of the other ones appearing
in the same amplitude ${\cal A}=<\prod_i{\cal O}_{\gamma_i}>$. This
is also a
novelty with
respect to the previous idea about this kind of topological
amplitudes,
that were expected to be only sensitive
to the spacetime manifold (like the so-called Donaldson invariants)
or that were explicitly shown to be constants (like the
gaugino condensates \cite{rossi}) and not step functions.
Consequently, the results of ref.s \cite{anomali,me} drastically
change our
vision of
four dimensional topological field theories. The topological map
mentioned in
the introduction
can push much further in this new direction.

Topological gravity, in the formulation proposed by Fr\`e and the
author
in ref.\ \cite{twist1}, was solved by the author in ref.\
\cite{anomali} with
the Eguchi-Hanson
gravitational instanton (and coupled to topological abelian
Yang-Mills theory).
This theory exhibits nonvanishing amplitudes associated to
3-dimensional closed
submanifolds.
Now, there is no nontrivial
3-cycle on the Eguchi-Hanson manifold, apart from one special case.
At the boundary of the moduli space,
the Eguchi-Hanson manifold degenerates to ${\IR^4/ \IZ_2}$. If the
3-cycle is
{\sl linked}
to the singular point of  ${\IR^4/\IZ_2}$, then the result is finite
and
nonzero, otherwise
the result is zero. The Eguchi-Hanson manifold has a noncontractible
2-sphere,
to which a topological observable is associated that also gives
nonvanishing amplitudes \cite{anomali}. Anyway,
it is interesting to notice that
the appearance of some concept of {\sl linking} seems to be quite a
general
feature
of four dimensional topological field theory.

The method proposed in section 2 of ref.\ \cite{anomali}
for solving topological field theories explicitly when the explicit
expression
of the instantons is known turns out to be more powerful than
expected.
Indeed, in certain cases, it is not necessary to know
the explicit expressions
of the instantons in order to compute their quantum
topological properties.
This fact will be exemplified by the theory solved in the next
section.
It is enough to have the explicit parametrization of a continuous
deformation
${\cal M}^\prime$
of the true moduli space ${\cal M}$.

Collecting the results of \cite{anomali,me}
and the present paper, the method of ref.\ \cite{anomali} has been
successful,
up to now,
for solving:

i) topological Yang-Mills theory with the BPST instanton;
this theory was also coupled to hyperinstantons \cite{hyper,ghyp}
in section 3 of ref.\ \cite{me};

ii) four dimensional topological gravity \cite{twist1}
with the Eguchi-Hanson instanton; this theory was also
coupled to topological abelian Yang-Mills theory;

iii) the topological version of the
two dimensional Ginzburg-Landau theory
of superconductivity, both on the plane and on the torus, with
and without external magnetic field.
There are no link invariants and the computation of the topological
amplitudes reduces to a straightforward, but tedious matter of
counting. It is somehow similar to what happens
in two dimensional topological gravity \cite{kontsevich}.

\section{The topological version of the Ginzburg-Landau theory
of superconductivity}
\label{uno}
\setcounter{equation}{0}

In this section I solve the topological version of
the Ginzburg-Landau theory
of superconductivity with $\lambda=1$.
Due to a theorem proved in ref.\ \cite{taubes2},
the result collects the quantum topological properties of the full
set of solutions to the classical field equations of the theory.
First, the theory is solved on $\IR^2$; in subsection \ref{due}
the results are generalized to the torus.

The Ginzburg-Landau theory of superconductivity is described by the
free-energy
\begin{equation}
{\cal L}=F_{\mu\nu}^2+{1\over 2}|D_\mu q^i|^2+{\lambda\over 8}{\cal
P}^2.
\label{lagra}
\end{equation}
where $F_{\mu\nu}={1\over 2}(\partial_\mu A_\nu-\partial_\nu A_\mu)$,
$Dq^i=dq^i+\varepsilon^{ij}Aq^j$ and ${\cal P}=1-q^2$.
$\lambda$ is the unique parameter that survives trivial redefinitions
and
distinguishes type I superconductors ($\lambda<1$) from
type II superconductors ($\lambda>1$). The value of $\lambda$ depends
on the
material,
on the amount of impurity and, very slightly, on the temperature.
$q$ is an order parameter that describes the distribution of Cooper
pairs.
For other details and the relation with the BCS microscopic theory
the reader
is referred
to ref.\ \cite{tinkham}.

The above ``free-energy'' will be called ``lagrangian''.
I take it as the lagrangian of a two dimensional quantum field
theory in the Euclidean framework and I consider the
associated functional integral.
In the intermediate situation $\lambda=1$, on which I focus from now
on
and that corresponds to the bosonic lagrangian
obtained by topologically twisting an N=2 supersymmetric
theory \cite{billofre}, ${\cal L}$
can be written as the sum of the squares of two instantonic
conditions plus a
topological
invariant
\begin{equation}
{\cal L}=\left(F_{\mu\nu}+{1\over 4}\varepsilon_{\mu\nu}{\cal P}
\right)^2+{1\over 4}\left(
D_\mu q^i-\varepsilon_{\mu\nu}D_\nu q^j\varepsilon_{ji}\right)^2
-{1 \over 2}\varepsilon_{\mu\nu}\Omega_{\mu\nu}.
\label{lag}
\end{equation}
The differential form $\Omega=\varepsilon_{ij}Dq^iDq^j+F{\cal
P}=\Omega_{\mu\nu}dx^\mu dx^\nu$ is closed, $d\Omega=0$, so that the
last term
of (\ref{lag})
is indeed a topological invariant, related to the so-called {\sl
fluxoid}.
One is interested in studying the {\sl vortex} equations
\begin{equation}
F_{\mu\nu}+{1\over 4}\varepsilon_{\mu\nu}{\cal P}=0,\quad\quad
D_\mu q^i-\varepsilon_{\mu\nu}D_\nu q^j\varepsilon_{ji}=0,
\label{equ}
\end{equation}
the second ones being the covariantized version of the Cauchy-Riemann
equations.

The quantization of the {\sl fluxoid}
corresponds to the Bohr-Sommerfeld quantization rule \cite{tinkham}.
Indeed, $\Omega$ is locally exact, $\Omega=d\omega$.
One finds $\omega=\varepsilon_{ij}q^idq^j+A{\cal P}=J+A$, $J=\rho v$
being the
electromagnetic current, $\rho=q^2$
being the density of Cooper pairs and $v$ being the velocity.
Now, the requirement of finite action implies,
in particular, that $\rho\rightarrow 1$ at infinity, so that
\begin{equation}
\int_{\IR^2}\Omega=-\oint_{C_\infty}\omega
=-\oint_{C_\infty}v+A=-\oint p\cdot dq=-2\pi n,
\end{equation}
having written $\partial\IR^2=-C_\infty$. It is clear that the
integer
$n$, that I call the {\sl vorticity}, can only take values with a
definite
sign.
The conventions have been chosen so that $n\geq 0$.
We have ${\cal L}=\pi n$, so that $n=0$ only with $q^2=1$, $A=$pure
gauge.

{}From the mathematical point of view, the integer number $n$ is the
{\sl
degree}
of the map $q:\IR^2\rightarrow \IR^2$ and is generalized to the case
$q:\Sigma_g\rightarrow M$ and gauge group $G$ according to
$\omega=i(\bar \partial {\cal K}-\partial {\cal K})+A^a{\cal P}_a$,
$\Sigma_g$ denoting a genus $g$ Riemann surface and $M$
denoting a K\"ahler target manifold with K\"ahler potential $\cal K$
and K\"ahler form $K=2i\partial\bar\partial {\cal K}\equiv
K_{ij}dq^idq^j$.
The vector potential $A^a$ gauges $M$-isometries associated with
Killing
vectors ${\bf k}_a=k_a^i(q){\partial\over \partial q^i}$, $[{\bf
k}_a,{\bf
k}_b]=- {f^c}_{ab}{\bf k}_c$.
${\cal P}_a$ is defined by $i_{{\bf k}_a}K=-d{\cal P}_a$, the
arbitrary
constants being fixed by imposing the equivariance condition
$0=K_{ij}k^i_a
k^j_b+{1\over 2}{f^c}_{ab}{\cal P}_c$.

The vortex equations admit two generalizations to four dimensions.
One is
simply their straightforward reinterpretation on four dimensional
K\"ahler
manifolds
and is related to N=1 supersymmetry, in the same way as in
two dimensions vortices are related to N=2 supersymmetry
\cite{billofre}.
The second type of equations, instead, were introduced by Fr\'e and
the author
in ref.\
\cite{hyper}.  Their solutions are called
{\sl hyperinstantons} and are related to N=2 supersymmetry.
They generalize the Cauchy-Riemann
equations to maps (called  {\sl tri-holomorphic maps}) between
hyperK\"ahler,
quaternionic K\"ahler, or simply almost quaternionic manifolds. They
can be
naturally
gauged and coupled to gravity. The fluxoid becomes the so-called
{\sl hyperinstanton number}.

Coming back to our problem, we note that the second equations of
(\ref{equ})
can be solved explicitly for the gauge field $A$ in terms of the
scalar
$q^i=|q|(\cos\theta,\sin\theta)$ as follows
\begin{equation}
A=d\theta-dx^\mu\varepsilon_{\mu\nu} \partial_\nu\ln |q|.
\label{solu}
\end{equation}
The angle $\theta$ is uniquely fixed only up to continuous
deformations,
that indeed correspond to the $U(1)$ gauge transformations.
In general, there are points $a_j$, $j=1,\ldots n$ such that
$d^2\theta=-2\pi\sum_{k=1}^n\delta(x-a_k)d^2x$ and one can choose
\begin{equation}
\theta=\sum_{k=1}^n{\rm arctg}{(x-a_k)_1\over
(x-a_k)_2}\equiv\sum_{k=1}^n\theta_k.
\label{theta}
\end{equation}
$\theta$ and $d\theta$ are singular in the points $a_k$. However, $A$
and $q^i$
must be regular everywhere. The only possibility for this to happen
is that
$|q|\rightarrow 0$ for $x\rightarrow a_k$.
Form (\ref{solu}) we see that in order for the singularity of
$d\theta$ to be
cancelled
by the one of $d\ln |q|$, $|q|$ has to behave like $|x-a_k|^{n_k}$
for
$x\rightarrow a_k$,
$n_k> 0$, so that $d^2\theta=-2\pi\sum_{k} n_k\delta(x-a_k)d^2x$.
 $n_k>0$ $\forall k$ means that there are only vortices and not
antivortices.
For the moment,
we assume $n_k=1$ $\forall k$,
the other cases taking place when some $a_j$'s coincide.

Inserting (\ref{solu}) into the first of (\ref{equ}) produces the
following Liouville equation for $\phi=\ln q^2$:
\begin{equation}
\Box\phi-4\pi\sum_{k=1}^n\delta(x-a_k)={\rm e}^\phi-1.
\label{lio}
\end{equation}
Although I am not going to
prove it rigorously, this equation
suggests that there
is one and only one solution
with $\phi\rightarrow 0$ at infinity
for any given set of $a_j$'s. The proof can be found
in ref.\ \cite{taubes}. We conclude that
the $a_j$'s are the moduli and
that the moduli space ${\cal M}_n$
is the symmetric product
$S^n\IR^{2}$ of $n$ copies of the plane.
In practice, we can integrate
each $a_j$ over $\IR^2$ and
divide the result by $n!$.
When some $a_j$'s coincide,
there should be a different
symmetry factor, but this does
not concern us, at least for now, because it only
affects an ${\cal M}_n$-subspace
of vanishing measure. I shall return later
to this point.

Thus, when $\lambda=1$ the vortices
can have arbitrary positions, without
interacting with one another. Instead,
when $\lambda<1$ they attract, while
when $\lambda>1$ they repel,
this being, qualitatively, the difference
between type I and type II superconductors \cite{tinkham}.

The BRST algebra of the theory is
\begin{equation}
sA_\mu=\partial_\mu C,\quad sC=0,\quad sq^i=-C\varepsilon^{ij}q^j.
\end{equation}
The BRST algebra of the topological version of the same
theory is obtained by introducing additional ghosts
$\psi_\mu$, $\phi$ and $\xi^i$ so as to kill any local degree of
freedom,
while preserving $s^2=0$:
\begin{eqnarray}
sA_\mu&=&\psi_\mu+\partial_\mu C,\quad sC=\phi,\quad sq^i
=\xi^i-C\varepsilon^{ij}q^j,\nonumber\\
s\psi_\mu&=&-\partial_\mu\phi,\quad\quad s\phi=0,\quad
s\xi^i=\phi\varepsilon^{ij}q^j-C\varepsilon^{ij}\xi^j.
\label{brs}
\end{eqnarray}
I shall solve the topological theory by using the method explained
in section 2 of ref.\ \cite{anomali} and by combining this method
with the theorems proved by Taubes in ref.s \cite{taubes,taubes2}
about the solutions of (\ref{equ}).

First of all, let us recall a very important fact proved in ref.\
\cite{taubes2}:
the set of solutions to the vortex equations (\ref{equ})
is the complete set of solutions to the field equations of the
theory.
In other words, not only it is true that the solutions to the
instanton
equations are solutions to the field equations, which is
obvious, but the converse is also true (with the boundary conditions
that
follow
from the requirement of finite action). This implies that we shall
find the quantum topological properties of the full set
of solutions to the field equations of the theory.

The second noticeable fact, already recalled in section \ref{qtp}, is
the
following:
the method of section 2 of
ref.\ \cite{anomali} is more powerful than expected, in the sense
that it is
not necessary to have the explicit parametrization of the space
${\cal M}$
of minima in order to find
the quantum topological properties of this space itself: it is enough
to
have the explicit parametrization of a subset ${\cal M}^\prime$
of the functional space that is a continuous deformation of the
true moduli space ${\cal M}$. In our case, I take
\begin{equation}
q^i=\prod_{k=1}^n{|x-a_k|\over \sqrt{\zeta+(x-a_k)^2}}
(\cos\theta,\sin\theta),
\label{solo}
\end{equation}
$\zeta$ being a useful extra parameter (but not a modulus).
The gauge field $A$, determined from (\ref{solu}),
and its field strength are
\begin{equation}
A=-\sum_{k=1}^n{\varepsilon_{\mu\nu}
(x-a_k)^\mu dx_\nu\over
\zeta +(x-a_k)^2},
\quad\quad
F=dA=-\zeta\sum_{k=1}^n{\varepsilon_{\mu\nu}dx^\mu dx^\nu\over
(\zeta+(x-a_k)^2)^2}.
\label{gfa}
\end{equation}
In this way, the covariantized Cauchy-Riemann equations are
satisfied.
What is not satisfied is the first equation of (\ref{equ}) or,
alternatively, the Liouville equation (\ref{lio}).
Nevertheless, the above configurations are, for any $\zeta>0$,
in one-to-one correspondence with the true solutions \cite{taubes}
and belong to the same topological sector:
\begin{equation}
-{1\over 2\pi}\int_{\IR^2}\Omega=-{1\over 2\pi}\int_{\IR^2}F=n.
\label{421}
\end{equation}
Consequently, I argue that the quantum topological properties are
also the same
and I solve the theory with (\ref{solo}) and (\ref{gfa}).

\let\picnaturalsize=N
\def\picsize{2.5in}
\ifx\nopictures Y\else{\ifx\epsfloaded Y\else\fi
\global\let\epsfloaded=Y
\centerline{\ifx\picnaturalsize N\epsfxsize \picsize\fi
\epsfbox{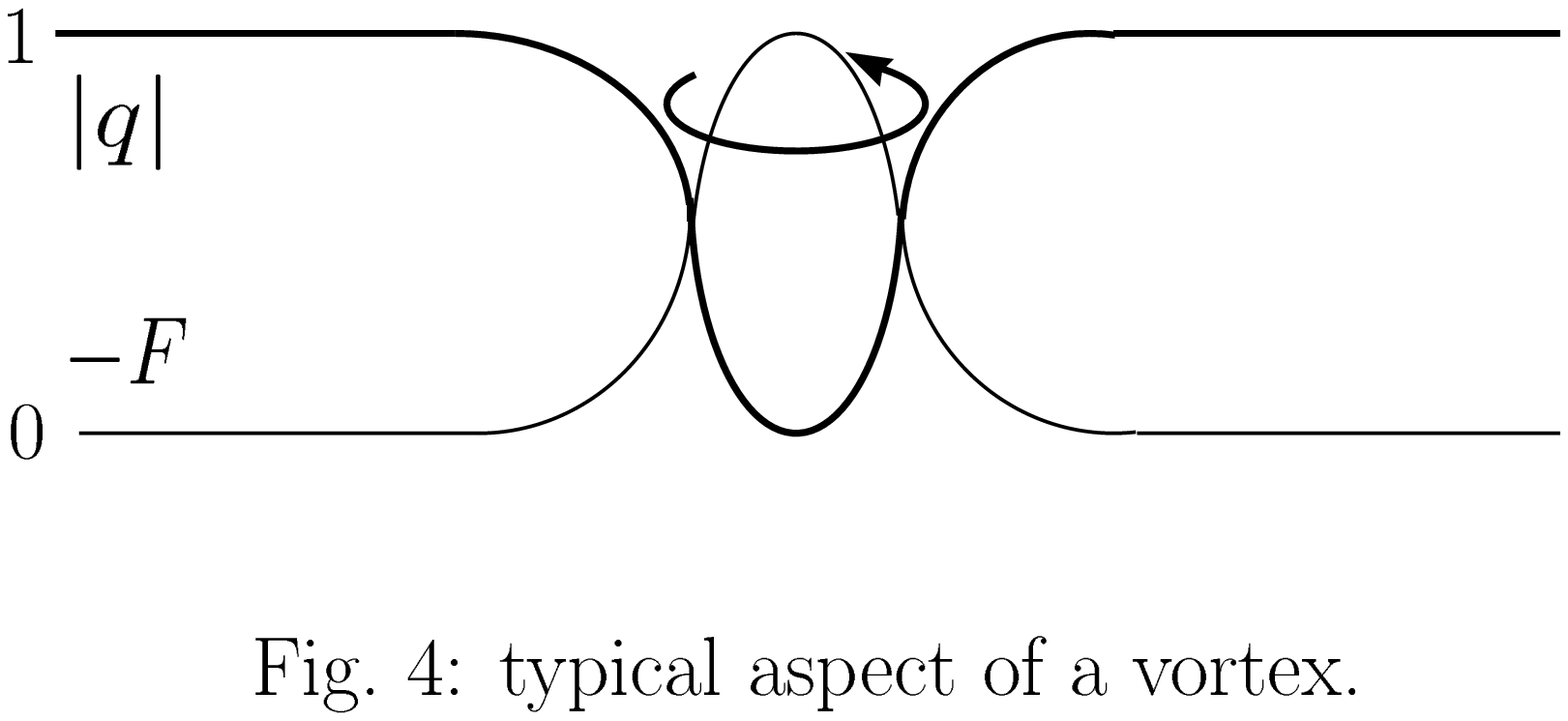}}}\fi

Notice that the ``wrong'' configurations that we use in replacement
of
the true solutions, nevertheless have the same basic physical
features
of the true solutions themselves. See Fig.\ 4. For example, $q$
vanishes
in the centers $a_k$ of the vortices and only there,
while the field strength is maximal in those points.

The key step in order to solve explicitly the topological theory is,
according to section 2 of ref.\ \cite{anomali}, the calculation of
the ghost
$C$, from which everything else follows automatically.
$C$ is fixed by a gauge-fixing the ``second'' gauge symmetry
(I use the terminology of \cite{anomali}), i.e.\ the symmetry
generated by
$\phi$.
Here, we do not possess a natural choice of this gauge-fixing
condition.
Moreover, we have used very simple explicit configurations
(that, nevertheless, could be the solutions of very complicated
equations)
in replacement of possibly complicated (and not explicitly known)
solutions of very simple instanton equations.
Very presumably the gauge-fixing that we need
is also very complicated. In brief, it is better to guess $C$
directly, without referring to any explicit gauge condition.
As before, we argue that the topological properties are independent
of all such
details, as long as we deal with well-defined quantities.

Noticing that every modulus is a translational one and
taking inspiration from
the solution of topological Yang-Mills theory with
the BPST instanton \cite{anomali},
we can guess
\begin{equation}
C=\sum_{k=1}^n{\varepsilon_{\mu\nu}
(x-a_k)^\mu da_k^\nu\over
\zeta +(x-a_k)^2},
\end{equation}
so that the BRST extensions of $\hat A=A+C$ and $\hat F=\hat d \hat
A=F+\psi+\phi$
of $A$ and $F$ (where $\hat d=d+s$ and
$s=\sum_{k=1}^nda_k^\mu{\partial\over
\partial a_k^\mu}$ is the ${\cal M}_n$-exterior derivative)
take the very simple forms
\begin{equation}
\hat A=-\sum_{k=1}^n{\varepsilon_{\mu\nu}
(x-a_k)^\mu d(x-a_k)^\nu\over
\zeta +(x-a_k)^2},\quad
\hat F=-\zeta\sum_{k=1}^n{\varepsilon_{\mu\nu}d(x-a_k)^\mu d(x
-a_k)^\nu\over
(\zeta+(x-a_k)^2)^2}.
\end{equation}
The choice of $C$ is not arbitrary, as it may seem at first sight
($C=0$, for example, is not good).
Indeed, while $A_\mu$, $q$ and $sA_\mu$, $sq$ are not normalizable,
in general
(and they do not need to be, since they are not physical fields),
$\psi_\mu$ and $\xi^i$ {\sl have} be normalizable, because they are
strict relatives of $F$ and $\Omega$ and so they appear in the
physical
observables, that are generated by the BRST extensions $\hat F$ and
$\hat
\Omega$.
This requirement guarantees that the physical amplitudes are
well-defined and topological.

The above choice of $C$ was studied in order to produce a
normalizable
$\psi_\mu$. A consistency-check is
that it also
produces a regular and normalizable $\xi^i$.
Indeed, (\ref{brs}) gives
\begin{eqnarray}
\xi=(\xi^1,\xi^2)&=&-\zeta\sum_{k=1}^n
{|q|\over |x-a_k|(\zeta+(x-a_k)^2)} \left(\sin
(\theta_k-\theta)da_k^1+
\cos(\theta_k-\theta)da_k^2,\right.\nonumber\\&&\left.
\cos (\theta-\theta_k)da_k^1+\sin(\theta-\theta_k)da_k^2 \right).
\label{kth}
\end{eqnarray}
This expression goes like ${1\over |x|^3}$ for
$|x|\rightarrow\infty$, so that
we only have to check that it is regular for $x\rightarrow a_k$.
Consider the
$k$-th term in (\ref{kth}):
when $x\rightarrow a_l$ with $l\neq k$, such term tends to zero, due
to the
factor $|q|$;
instead when $x\rightarrow a_k$, ${|q|\over |x-a_k|}$ is regular and
$\theta-\theta_k$ has a well-defined limit (note that $\theta_k$ has
no
well-defined limit
for $x\rightarrow a_k$).

The observables of the theory are
\begin{equation}
{\cal O}^m(x)={(-1)^m\over (2\pi)^m}\phi^m(x),
\quad
{\cal O}^{(d)}={(-1)^d\over (2\pi)^d}\int_{\IR^2}\hat F^d,
\label{ob}
\end{equation}
plus the ones generated in a completely similar way
by the BRST extension $\hat \Omega$ of $\Omega$.
Here we focus on the observables (\ref{ob}), because (\ref{421})
suggests that
$\hat \Omega\sim\hat F$ inside the amplitudes.
This fact has been explicitly checked in the simplest cases.
Moreover, the calculations with $\hat F$ are much simpler than the
ones with
$\hat \Omega$. Indeed, recalling that (\ref{solo}) and (\ref{gfa})
are a continuous deformation of the true solutions for {\sl any}
value of
$\zeta$, we can choose the most convenient $\zeta$, that is
$\zeta=0$.
In this limit we get only delta functions
\begin{equation}
-{1\over 2\pi}\hat F\rightarrow
\sum_{k=1}^n\delta(x-a_k)\,d^2(x-a_k),
\label{323}
\end{equation}
so that the calculation of the topological amplitudes reduces to a
pure
combinatorial
counting, while any integration is trivial. This is more or less what
happens in two dimensional topological gravity, where Strebel's
theorem
\cite{strebel}
allows one to reduce
the problem to a simple combinatorial counting, that can also be
encoded into
a matrix model \cite{kontsevich}. There, the limit $\zeta\rightarrow
0$
corresponds
to deforming the Riemann surface so as to take the punctures at
infinity and
reduce basically to the case in which the Riemann curvature has a
delta support on the punctures.

The situation is different, instead, in the case of topological
Yang-Mills
theory with the BPST instanton, as I recalled in section \ref{qtp},
where
not every observable inside an amplitude can be reduced to a delta
function,
but one,
and only one (the result being independent
of  {\sl which } one), has to be replaced by the Chern-Simons form,
this
being the simple reason why link numbers appear instead of constant
amplitudes.

The generic amplitude that we consider is
\begin{equation}
{\cal A}_{\{d\}m}=<{\cal O}^m(x)\cdot \prod_{i=1}^k{\cal O}^{(d_i)}>.
\label{ampl}
\end{equation}
The vorticity contributing to this amplitude is
\begin{equation}
n=\sum_{i=1}^k(d_i-1)+m.
\end{equation}
 (\ref{323}) shows immediately that  ${\cal A}$ is independent of the
positions
of the local observables. This is why in (\ref{ampl}) I have put the
local observables in the same point $x$.
The computation of the above amplitudes is now a simple, but tedious
matter of
counting.
One has to decompose
\begin{equation}
d^2(x-a_i)=d^2x+d^2a_i-
\varepsilon_{\mu\nu}dx^\mu da_i^\nu,
\label{diff}
\end{equation}
Were it not for the last term, to which we shall refer as
the {\sl double product} in the sequel,
it would be immediate to extract
the relevant components to the various observables.
Let us write
\begin{equation}
{\cal A}_{\{d\}m}={1\over n!}\sum_{j=0}^k C^{(j)}_{\{d\}m},
\end{equation}
$C^{(0)}_{\{d\}m}$ denoting the contribution in which the
double products are completely neglected and
$C^{(j)}_{\{d\}m}$ being the contribution where $j$ and only $j$
pairs of
double products are taken into account.

I show in the appendix that
\begin{equation}
C^{(0)}_{\{d\}m}=n!\prod_{i=1}^k d_i(n-d_i+1),
\quad\quad C^{(j)}_{\{d\}m}=-(j-1)\,C^{(0)}_{\{d\}m}
\sum_{\sigma_j}\,\,\prod_{p\in\sigma_j}
{(d_p-1)\over (n-d_p+1)},
\label{prima}
\end{equation}
$\sigma_j$ denoting a $j$-uple of elements of the set $\{1,\ldots
k\}$.
The final sum gives
\begin{equation}
{\cal A}_{\{d\}m}=n^{k-1}\, m\prod_{i=1}^kd_i.
\label{dopo}
\end{equation}
Indeed, it is easy to show that $\forall n$, given $k$ integer
numbers
$d_1,\ldots d_k$, the following identity holds
\begin{equation}
f_k\equiv\sum_{j=0}^k\sum_{\sigma_j}\,\,\prod_{p\in\sigma_j}
(d_p-1)\,\,\prod_{q\not\in\sigma_j} (n-d_q+1)=n^k.
\end{equation}
This is straightforward, by induction. For $k=0$ we have $1=1$. For
generic
$k$,
distinguishing those $\sigma_j$ that contain $k$ from those
that do not contain it, we can write
\begin{eqnarray}
f_k&=&
(d_k-1)\sum_{j=1}^k\sum_{\sigma_{j-1}^\prime}
\prod_{p\in\sigma_{j-1}^\prime}
(d_p-1)\prod_{q\not\in\sigma_{j-1}^\prime} (n-d_q+1)\nonumber\\&&
+(n-d_k+1)\sum_{j=0}^{k-1}\sum_{\sigma_j^\prime}\,
\prod_{p\in\sigma_j^\prime}
(d_p-1)\prod_{q\not\in\sigma_j^\prime} (n-d_q+1)=n\,f_{k-1},
\end{eqnarray}
where $\sigma_j^\prime$ are $j$-uples of elements of the set
$\{1,\ldots
k-1\}$.
In a similar way one proves that
\begin{eqnarray}
g_k&\equiv&\sum_{j=0}^kj\sum_{\sigma_j}\prod_{p\in\sigma_j}
(d_p-1)\prod_{q\not\in\sigma_j} (n-d_q+1)=(d_k-1)n^{k-1}
\nonumber\\&&+n
\sum_{j=0}^{k-1}j\sum_{\sigma_j^\prime}\prod_{p\in\sigma_j^\prime}
(d_p-1)\prod_{q\not\in\sigma_j^\prime} (n-d_q+1)=
ng_{k-1}+n^{k-1}(d_k-1).
\end{eqnarray}
We have ${\cal A}_{\{d\}m}=h_k\prod_{i=1}^kd_i$, with
$h_k=f_k-g_k=n^k-g_k=nh_{k-1}-n^{k-1}(d_k-1)$.
With $h_0=1$ it is easy to prove that the recursion relation is
solved by
$h_k=n^{k-1}\left[n-\sum_{i=1}^k(d_i-1)\right]=n^{k-1}m$,
from which the result (\ref{dopo}) follows.

A check that formula (\ref{dopo})
is correct is that it turns out to be proportional to $m$, although
no
$C^{(j)}_{\{d\}m}$ is. Indeed, ${\cal A}_{\{d\}0}$ has to be zero,
for the following simple reason.
Expression (\ref{323}) shows that the arguments of the delta
functions are {\sl
differences}
of points. When $m=0$, the local observable is absent and there are
$n$
delta functions depending on $n-1$ {\sl differences}.
Consequently, there would be a $\delta(0)$,
which, however, is multiplied by a zero coefficient.
Instead, when $m\neq 0$, there is one additional point around,
namely the point $x$ where the local
observable ${\cal O}^m(x)$ is placed, so
that there  are $n$ delta functions for $n$ differences.

Moreover, ${\cal A}_{\{d\}m}$ is proportional to any $d_i$,
consistently with
the fact
that ${\cal O}^{(d)}=0$ for $d=0$.

\subsection{The $\tau$-function}

With the result (\ref{dopo}) one can compute the $\tau$-function
\begin{equation}
\tau[t]=\int d\mu\,\,{\rm exp}\left({\sum_{i=0}^\infty t_i{\cal
O}^{(i)}}
\right),
\label{333}
\end{equation}
${\cal O}^{(0)}$ denoting the local observable ${\cal O}(x)$. One
finds
\begin{equation}
\tau[t]=\sum_{n=0}^\infty\,{\rm e}^{nt_1}\tau_n[t],\quad
\tau_0[t]=1,\quad
\tau_n[t]={1\over n}\sum_{{\cal K}_n}
{t_0^{k_0}\over (k_0-1)!}\prod_{i=2}^\infty {(int_i)^{k_i}\over k_i!}
\,\,{\rm for}\,n>0,
\end{equation}
${\cal K}_n$ denoting a string of natural numbers
$\{k_0,k_2,k_3,\ldots\}$ such that $k_0+\sum_{i=2}^\infty
(i-1)k_i=n$,
$k_0>0$ and $k_i\geq 0$ $\forall i\geq 1$. $\tau_n[t]$ is a finite
polynomial
$\forall n$.

Putting $t_i=0$ for $i>1$ in (\ref{333}) one has
\begin{equation}
\tau[t_0,t_1]={\rm exp}\left(t_0\,{\rm e}^{t_1}\right).
\end{equation}
Putting $t_i=0$ for $i>2$, instead, the $\tau$-function
is convergent for $|2{\rm e}t_2{\rm e}^{t_1}|<1$:
\begin{equation}
\tau[t_0,t_1,t_2]=t_0\sum_{n=0}^\infty{1\over n!}\,{\rm
e}^{nt_1}(t_0+2nt_2)^{n-1},
\end{equation}

The first $\tau_n[t]$'s are
\begin{eqnarray}
\tau_1[t]&=&t_0,\quad\quad\tau_2[t]={1\over
2}t_0^2+2t_0t_2,\quad\quad
\tau_3[t]={1\over 6}t_0^3+2t_0^2t_2+6t_0t_2^2+3t_0t_3,\nonumber\\
\tau_4[t]&=&{1\over 24}t_0^4+t_0^3t_2+8t_0^2t_2^2+3t_0^2t_3+{64\over
3}t_0t_2^3+24t_0t_2t_3+4t_0t_4,\nonumber\\
\tau_n[t]&=&{1\over n!}t_0^n+{2\over (n-2)!}t_0^{n-1}t_2+{1\over
(n-3)!}t_0^{n-2}(2nt_2^2+3t_3)
\nonumber\\&&+
{1\over (n-4)!}t_0^{n-3}\left({4\over 3}n^2
t_2^3+6nt_2t_3+4t_4\right)+\cdots.
\end{eqnarray}
Summing term by term, we get expressions like
\begin{eqnarray}
\tau[t]&=&(1+2t_0t_2+3t_0t_3+4t_0t_4+\cdots+6t_0t_2^2+24t_0t_2t_3+
\cdots)\,{\rm
exp}\left(t_0\,{\rm e}^{t_1}\right)\nonumber\\
&&+(2t_0^2t_2^2+6t_0^2t_2t_3+\cdots)\,{\rm exp}\left(t_1+t_0\,{\rm
e}^{t_1}\right)+
\cdots.
\end{eqnarray}

\subsection{Impurities}

Let us now suppose that, for some reason\footnotemark
\footnotetext{In the realm of topological field theory,
such situation can perhaps be obtained
{\sl via} a constraining
mechanism like the one studied in
ref.\ \cite{constr} (constrained topological field
theory).},
the integral over the moduli space ${\cal M}$
is repalced by the integral over some proper
${\cal M}$-subspace ${\cal M}^{\{n\}}$, for example a diagonal
subspace, i.e.\ a subset where the positions
of certain vortices coincide, so that
\begin{equation}
-{1\over 2\pi}\hat F=
\sum_{j=1}^\infty j\sum_{i=1}^{n_j}
\delta(x-a^i_j)\, d^2(x-a^i_j).
\end{equation}
This means that there are $n_j$ vortices with
vorticity $j$. The total vorticity is $v=
\sum_{j=1}^\infty j n_j$.
The (constrained) moduli space is
${\cal M}_v^{\{n\}}=\bigotimes_{j=1}^\infty
S^{n_j}\IR^2$, $n={\rm dim}\,
{\cal M}_v^{\{n\}}=\sum_{j=1}^\infty n_j$.
The symmetry
factor is $\prod_{j=1}^\infty n_j!$. Because
of the reduced symmetry,
we speak about ``impurity'', when referring to the projection onto
${\cal M}_v^{\{n\}}$,
the kind of impurity being specified by the
set of integer numbers $\{n\}=\{n_1,\ldots n_j,\ldots\}$.
The amplitudes are formally the same
as in (\ref{ampl}) and I denote them
by ${\cal A}_{\{d\}m}^{\{n\}}$.

In the
general case, the counting is much more involved
than before. One has to go through the detailed derivation of the
appendix,
where the case without impurities is analysed in full detail,
and improve the arguments when necessary.

\let\picnaturalsize=N
\def\picsize{3.0in}
\ifx\nopictures Y\else{\ifx\epsfloaded Y\else \fi
\global\let\epsfloaded=Y
\centerline{\ifx\picnaturalsize N\epsfxsize \picsize\fi
\epsfbox{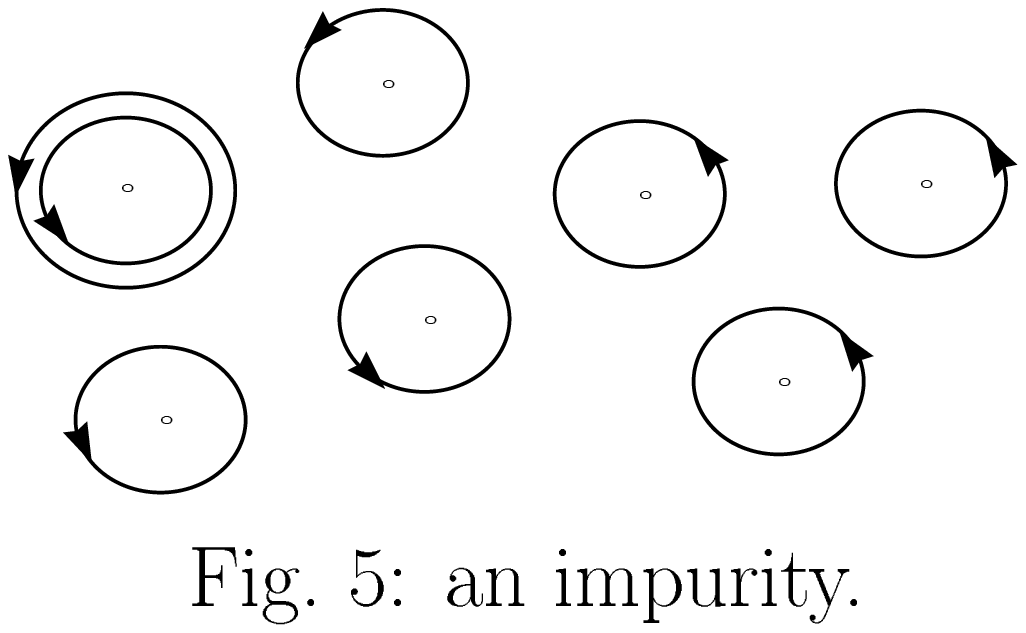}}}\fi

For example, in the case with
one impurity, say when the
vortex placed in $a_1$ has vorticity $I$, one finds
\begin{eqnarray}
{\cal A}_{\{d\}m}^{(I)}&=&
I\prod_{i=1}^kd_i(n-d_i+I)
\sum_{j=0}^k(1-j)
\sum_{\sigma_j}\prod_{p\in\sigma_j}
{d_p-1\over n-d_p+I}\left[n+(I-1)\left(j-
\sum_{q\not\in\sigma_j}
{d_q-1\over n-d_q+I}\right)\right]\nonumber\\
&=&Iv^km\prod_{i=1}^kd_i,
\end{eqnarray}
where $n=m+\sum_{i=1}^k(d_i-1)$, as before, while $v=n+I-1$.
As we see, the various contributions still sum up nicely as before.
Assuming
that this also holds in the general case, a general formula that
agrees with
the results found so far is
\begin{equation}
{\cal A}^{\{n\}}_{\{d\}m}=<{\cal O}^m(x)\cdot \prod_{i=1}^k{\cal
O}^{(d_i)}>=
v^k (n-1)!\,m\prod_{i=1}^kd_i\prod_{j=1}^\infty{j^{n_j}\over n_j!}.
\label{final}
\end{equation}

The statistical impurity distribution is the Bose-Einstein one, where
the role
of the energy is played by $v$. More precisely, I denote  the energy
of a
vortex unity
by $\varepsilon$, so that the
total energy is  $E=\varepsilon v$.
A good parameter that quantifies the amount of impurity is
\begin{equation}
d\equiv {\bar v-\bar n\over \bar n}\geq 0.
\end{equation}
A small $d$ corresponds to a small amount of dirtiness.
It is easy to check that lowering the temperature
reduces the number of impurities: $d={1\over \bar n+1}\,{\rm
e}^{-\beta\varepsilon}+
{\cal O}({\rm e}^{-2\beta\varepsilon})$ for $\beta\rightarrow
\infty$.
The standard Bose statistical distribution partially answers the
question
raised in point iii) of the introduction. It is restricted,
however,  to the classical topological invariant
$v$ only. It should be possible, using (\ref{final}), to define a
generalized
statistical distribution for the quantum topological sectors ${\cal
Q}$.

To conclude this subsection, let us note
that a straightforward four dimensional generalization
of the counting problem that we had to deal with is
obtained by considering observables like
\begin{equation}
{\cal O}^{(d)}=\int_{\IR^4}\hat{\cal Q}^d,\quad {\cal O}^m(x)=
\left[\sum_{k=1}^nn_k\delta(x-a_k)\,d^4a_k\right]^m,\quad
\hat{\cal Q}=\sum_{k=1}^nn_k\delta(x-a_k)\,d^4(x-a_k).
\end{equation}
The amplitudes are formally the same as in
(\ref{ampl}), although their evaluation
appears to be less straightforward.
This problem could be related to some kind of hyperinstantons
\cite{hyper,ghyp}.

\subsection{Case of the torus and presence of an external magnetic
field}
\label{due}

With a simple recursion relation,
the results of the previous section can be extended to the case of
the
torus and to the presence of an external magnetic field.
The existence theorems for solutions to the vortex equations
on compact Rieman surfaces can be found in ref.\ \cite{prada}.

Let $V$ denote the volume of the torus. Despite the fact that the
volume is not a topological quantity, it is relevant to our problem.
Indeed, integrating the first equation of (\ref{equ})
on a compact Riemann surface $\Sigma$, one gets
\cite{prada} in our units
\begin{equation}
v\leq \left[{V\over 4\pi}\right]\equiv N.
\label{bound}
\end{equation}
This is an interesting example in which a
non-topological quantity enters a topological field theory:
still, any correlation function is topological,
but the {\sl range} of the classical topological invariant
depends on the size of $\Sigma$. The $\tau$-function
becomes a finite sum.

On the torus, besides the pointlike vortices that we have on the
plane,
there can be a vorticity around the handle, corresponding to the
constant
two-form
\begin{equation}
-{1\over 2\pi}F={p\over V}\,d^2x, \quad\quad\quad p\in \IZ.
\end{equation}
This extra contribution does not change the moduli space.
So, we are lead to consider observables generated by
\begin{equation}
-{1\over 2\pi}\hat F^\prime=
\sum_{j=1}^\infty j\sum_{i=1}^{n_j}
\delta(x-a^i_j)\, d^2(x-a^i_j)+{p\over V}\,d^2x.
\label{expr}
\end{equation}
The prime is used to denote genus one quantities.
Now, the total vorticity is $p+v=p+\sum_{j=1}^\infty jn_j$, while
the dimension of the moduli space $\otimes_{j=1}^\infty
S^{n_j}\Sigma$ is $n=\sum_{j=1}^\infty n_j$,
as before. The symmetry factor is also unchanged.

A glance at the correlation functions shows that they are independent
of $V$, as it must be, although $V$ enters (\ref{expr}) explicitly.
In
practice,
one can replace (\ref{expr}) with a more convenient expression
obtained with the substitution ${1\over V}\rightarrow\delta(x-y)$,
$y$ being an arbitrary point of $\Sigma$. One can write
\begin{equation}
{{\cal O}^\prime}^m(x)={\cal O}^m(x),\quad\quad
{{\cal O}^\prime}^{(d)}=p\,d\,{\cal O}^{d-1}(y)+{\cal O}^{(d)}.
\end{equation}
Thus, provided (\ref{bound}) holds, we have, using (\ref{final}),
\begin{eqnarray}
{\cal A}_{\{d\}m}^{\{n\}p}&=&<{{\cal O}^\prime}^m(x)
\cdot \prod_{i=1}^k {{\cal O}^\prime}^{(d_i)}>=
(n-1)!\prod_{i=1}^kd_i\prod_{j=1}^\infty{j^{n_j}\over n_j!}
\sum_{j=0}^kp^{k-j}v^j\sum_{\sigma_j}\left(n-\sum_{i\in\sigma_j}
(d_i-1)\right)\nonumber\\
&=&(np+mv)\,(p+v)^{k-1}
(n-1)!\prod_{i=1}^kd_i\prod_{j=1}^\infty{j^{n_j}\over
n_j!}.
\end{eqnarray}
The inclusion of the Abelian differentials of the torus can be
achieved by
replacing $A$ with $A+u_\mu dx^\mu$. The vector $u_\mu$ is a modulus
and
belongs to the fundamental cell of the reciprocal lattice of the
torus.
$F$ is unchanged, but
$\hat F$ gets the extra contribution $du_\mu dx^\mu$.
In this case, one can also construct observables by integrating over
the cycles of the torus. However, the correlation functions reduce
to the known ones.

Taking $V\rightarrow\infty$, the bound (\ref{bound}) drops out
and $p$ is no longer required to be an integer. Then one
recovers the case of $\IR^2$ in presence of an external
magnetic field $p$. For large $p$,
the amplitude behaves like $p^k$. So, for $k$ even there is a minimum
when the magnetic field has certain fractional values, precisely when
\begin{equation}
p=-{v\,(n+m\,(k-1))\over n\,k}.
\end{equation}
It is curious to note that some fractional values of the
external magnetic field play
a special role (also note the condition $k$=even), although
they do not seem to be related to the values observed
in the fractional quantum Hall effect.

\subsection{Perspectives}

Assuming that (\ref{lagra}) is the complete free-energy (as a matter
of fact,
it only contains the first few terms of an expansion in $q$ and its
derivatives
\cite{tinkham}),
the general remarks made in the first part of this paper and the
results found
in the present section should stimulate the curiosity of seeing what
happens
experimentally
in the special case $\lambda=1$, where the classical theory fails and
the
functional integral is required.
Such situations have to be treated {\sl via} the topological
embedding, so that
the quantum topological invariants computed in
this section should have a relevant counterpart
in the experimental results.
When including the higher order corrections to the free energy,
it can happen that the minima become discrete,
nevertheless the consequences of the
topological embedding should still be visible,
since
(\ref{lagra}) is a good approximation.
Experimentally,
one should manage
to tune $\lambda$ across
the critical value and examine
what sort of transition occurs.
The general theoretical set-up
developed here suggests
that one should find a
qualitatively new kind of transition.

\section{Conclusion}
\label{concl}
\setcounter{equation}{0}

The best feature of the topological embedding is that it is
{\sl testable} in nature and this possibility does not seem
so distant. For example, one should estabilish
whether QCD or QCD$^*$ better describe the real world.
A key qualitative feature of the topological embedding is that any
event is
placed in
a {\sl single} topological sector.
However, the new idea is quite general and can be fruitfully applied
to many other problems, for example superconductivity in a very
special case.
This fact makes the new approach more powerful and more easily
testable.
In the present paper the fundamental guidelines in this research area
were estabilished.
Surely it is worth insisting in this direction.

\vskip .3in
\begin{center}
{\bf Acknowledgements}
\end{center}

I am especially grateful to R.\ Iengo for an interesting discussion
in which he strongly
defended the ``orthodox'' point of view and stimulated indirectly
some of the ideas contained in subsection \ref{norm}.
I am also indebted to P.\ Fr\`e for drawing my attention
to question iii) of the introduction and to A.\ Johansen
for drawing my attention to ref.\ \cite{prada}.
Finally, I would like to thank M.\ Martellini, A.S.\ Cattaneo,
G.C.\ Rossi, M.\ Bianchi and F.\ Fucito for valuable discussions.
This research was supported in part by the
Packard Foundation and by NSF grant PHY-92-18167.

\section{Appendix: proof of formul\ae\ (4.22)}
\label{proof}
\setcounter{equation}{0}

In this appendix, I prove formul\ae\ (\ref{prima}).
The proof is done in several steps. It is one of those cases in which
it
can be easier to work out the proof independently than reading it.
Anyway, for completeness, I have to write down the details.

\underline{\sl Step 1:} $C^{(0)}_{\{d\}m}$.

Let us begin by computing $C^{(0)}_{\{d\}m}$.
Each observable involves the polynomial $\hat F$, that is a sum of
monomials of the form $\delta(x-a)\,d^2(x-a)$, raised to the power
$m$ (local observable) or $d_i$ (nonlocal). When expanding the
polynomial,
each monomial can only be raised to the powers 0 or 1, because the
square of $d^2(x-a)$ vanishes. So, there is a combinatorial factor
\begin{equation}
m!\prod_{i=1}^kd_i!
\end{equation}
simply coming from the expansion of the polynomials.
Now, one has distribute the powers 0 and 1
and count the number of possibilities.
The local observable ${\cal O}^m(x)$, for which $d^2(x-a)$ reduces to
$d^2a$,
has to saturate the integrations over $m$ moduli.
There are $\left(\matrix {n\cr m}\right)$ ways of achieving this.
So, we can multiply by $\left(\matrix {n\cr m}\right)$ and assume
that, say,
the first $m$ moduli-integrations are saturated. Up to now, we have
\begin{equation}
n!\prod_{i=1}^kd_i!{1\over (n-m)!}.
\label{fac1}
\end{equation}
At this point, we have to distribute the integrations coming from the
first
nonlocal
observable ${\cal O}^{(d_1)}$. There are $d_1-1$ moduli-integrations
and one spacetime-integration (such integration being part of the
observable
and
not of the amplitude). The spacetime-integration is naturally
associated to a
certain
modulus $a$, since one only has differentials like $d^2(x-a)$.
Two situations can happen:

i) if the spacetime-integration is associated to one of the first $m$
moduli
$a_1,\ldots a_m$
($m$ possibilities), then the $d_1-1$ moduli integrations can be
chosen
in $\left(\matrix {n-m\cr d_1-1}\right)$ ways;

ii) if the spacetime-integration is associated to any other modulus
$\bar
a=a_i$,
$i>m$ ($n-m$
possibilities), then the $d_1-1$ moduli integrations can be chosen
in $\left(\matrix {n-m-1\cr d_1-1}\right)$ ways (the $\bar
a$-integration
cannot be included in this case).

The factor due to ${\cal O}^{(d_1)}$ is thus
\begin{equation}
m \left(\matrix {n-m\cr d_1-1}\right)+(n-m)
\left(\matrix {n-m-1\cr d_1-1}\right)=(n-d_1+1)
\left(\matrix {n-m\cr d_1-1}\right),
\end{equation}
so that, together with (\ref{fac1}), we have, so far,
\begin{equation}
n!\,d_1(n-d_1+1){1\over (n-m-d_1+1)!}\prod_{i=2}^k d_i!.
\end{equation}
and we can assume that the first $m+d_1-1$ moduli integrations are
saturated.
Now, for ${\cal O}^{(d_2)}$, one can proceed exactly as before,
with $m\rightarrow m+d_1-1$, and so on. The final result is the
claimed one.

\underline{\sl Step 2:} $C^{(1)}_{\{d\}m}$.

$C^{(1)}_{\{d\}m}$ is trivially zero. Indeed, there is no possibility
with
only one pair of double products. We can assume that the nonlocal
observable
interested in this pair of double products is ${\cal O}^{(d_1)}$.
We call it the {\sl special observable}.
Each double product is associated with a moduli integration. Let us
say that
the ${\cal O}^{(d_1)}$-pair of double
products is associated to $\bar a$ and $\bar b$, with
$\bar a\neq \bar b$. That means that the $\bar a$ and $\bar b$
integrations
cannot be completely saturated, unless some other double products,
coming from other nonlocal observables, join the game. This cannot
happen for $C^{(1)}_{\{d\}m}$ by assumption, but happens in the other
cases.

\underline{\sl Step 3:} $C^{(2)}_{\{d\}m}$.

In this case, instead, the $\bar a$ and $\bar b$ integrations can be
completely
saturated,
because there are two nonlocal special observables.
There will be a sum $\sum_{\sigma_2}$ over the
set of couples $\sigma_2$ of the special observables. So, we can
restrict to
$\sigma_2=\{1,2\}$.
We can assume that the pairs of double products in ${\cal O}^{(d_1)}$
and ${\cal O}^{(d_2)}$
correspond to $a_{m+1}$ and $a_{m+2}$, this producing a factor
$\left(\matrix {n-m\cr 2}\right)$. The remaining $d_1-2$ integrations
of ${\cal O}^{(d_1)}$
can be chosen to saturate $a_{m+3},\ldots a_{m+d_1}$, this producing
a factor $\left(\matrix {n-m-2\cr d_1-2}\right)$.
Finally, the remaining $d_2-2$ integrations
of ${\cal O}^{(d_2)}$
can be chosen to saturate $a_{m+d_1+1},\ldots a_{m+d_1+d_2-2}$, this
producing
a factor $\left(\matrix {n-m-d_1\cr d_2-2}\right)$.
Taking into account that
\begin{equation}
(-\varepsilon_{\mu\nu}dx_1^\mu d\bar a^\nu)
(-\varepsilon_{\rho\sigma}dx_1^\rho d\bar b^\sigma)
(-\varepsilon_{\alpha\beta}dx_2^\alpha d\bar a^\beta)
(-\varepsilon_{\gamma\delta}dx_2^\gamma d\bar b^\delta)=
-2\,d^2x_1\, d^2x_2\,d^2\bar a\, d^2\bar b,
\label{path}
\end{equation}
we see that there is a further factor $-2$ with respect to before.
Collecting the factors computed so far, we have
\begin{equation}
(-2)n!\prod_{i=1}^kd_i!{1\over (n-m)!}\sum_{\sigma_2=\{i,j\}}
\left(\matrix {n-m\cr 2}\right)
\left(\matrix {n-m-2\cr d_i-2}\right)
\left(\matrix {n-m-d_i\cr d_j-2}\right)(\cdots),
\end{equation}
the dots standing for the factors that remain to be computed.
{}From this point onwards one can proceed, for any $i$ and $j$, as in
Step 1,
assuming that the first $m+d_i+d_j-2$ moduli integrations are
saturated.
So, one gets
\begin{equation}
-n!\prod_{i=1}^kd_i!\sum_{\sigma_2=\{i,j\}}
{1\over (d_i-2)!(d_j-2)!}\prod_{k\neq i,j}{n-d_k+1\over (d_k-1)!},
\end{equation}
as desired.

\underline{\sl Step 4:} $C^{(j)}_{\{d\}m}$, $j>2$.

The case $j=2$ is sufficient to illustrate, with some straightforward
adaptations,
what happens in general. When $j>2$, one has many more possible ways
of
rearranging
the pairs of double products.
Of course there is a sum $\sum_{\sigma_j}$ and
we focus on $\sigma_j=\{1,\ldots j\}$, which means that the special
observables
are ${\cal O}^{(d_1)},\ldots {\cal O}^{(d_j)}$.
One can distinguish some closed paths, in the following way.
We have to consider a $j\times j$ matrix.
In each row, two and only two entries are 1, all the other ones being
0.
The same for each column.
The 1's correspond to the positions of the double products. The rows
represent
different moduli, while the columns represent different special
observables.
One can always exchange the rows and the columns in such a way
that a certain number of closed paths is obtained. For example, for
$j=6$
one can have situations like
\begin{equation}
\left[\matrix{
1&1&&&&\cr 1&&1&&&\cr &1&&1&&\cr &&1&&1&\cr &&&1&&1\cr
&&&&1&1}\right]
\left[\matrix{
1&1&&&\cr 1&1&&&\cr &&1&1&&\cr &&1&&1&\cr &&&1&&1\cr
&&&&1&1}\right]
\left[\matrix{
1&1&&&\cr 1&&1&&\cr &1&1&&&\cr &&&1&1&\cr &&&1&&1\cr
&&&&1&1}\right]
\left[\matrix{
1&1&&&\cr 1&1&&&\cr &&1&1&&\cr &&1&1&&\cr &&&&1&1\cr
&&&&1&1}\right]
\label{matri}
\end{equation}
In the first case there is only one closed path, in the second and
third cases
there are two,
in the last case there are three. In general,
the number of such closed paths is between 1 and the integral part of
$j/2$.
A concrete example of a closed path is given in eq.\ (\ref{path}),
which
represents
$\left(\matrix{1&1\cr 1&1}\right)$.
One has to count the number of possibilities like the ones
illustrated above,
with the appropriate weight.
Any closed path has a factor $-2$, like in (\ref{path}).
Of course, there is the overall factor (\ref{fac1}).

For convenience, let us define
\begin{equation}
a_j=n!\prod_{i=1}^kd_i!\sum_{\sigma_j}\prod_{p\in\sigma_j}{1\over
(d_i-2)!}
\prod_{q\not\in\sigma_j}{(n-d_q+1)\over (d_q-1)!}.
\end{equation}

{\it a)} The case with one and only one closed path has a weight
$-a_j(j-1)!$.
Refer to the first example of (\ref{matri}).
There is a factor $-2$ for the closed path.
The 1's in the first column can be fixed in
$\left(\matrix{n-m\cr 2}\right)$ ways. Consider the first row.
I can fix the position of the second 1 in $j-1$ ways. With a factor
$j-1$,
I can put it in position $(1,2)$. Now, consider the second column.
I cannot put a 1 in position $(2,2)$, since it would close a path.
This is what
happens in
the second case of (\ref{matri}) and will be discussed below.
With a factor $n-m-2$, I put the 1 in position
$(3,2)$. Then, consider the second row. I can fix the position of the
second
1 in $(2,3)$, gaining a factor $j-2$. Proceeding in this way, one
obtains the
weight
$-(j-1)!{(n-m)!\over (n-m-j)!}$ after distributing the double
products.
Next, the $d_i-2$ remaining moduli integrations, $i=1,\ldots j$, give
factors
\begin{equation}
\left(\matrix{n-m-j\cr d_1-2}\right)\left(\matrix{n-m-j-d_1+2\cr
d_2-2}\right)\cdots
\left(\matrix{n-m-j-\sum_{i=1}^{j-1}(d_i-2)\cr d_j-2}\right).
\end{equation}
For the other observables ${\cal O}^{(l)}$, $l=j+1,\ldots k$, one can
proceed
as in the second part of Step 1, with $m\rightarrow m+\sum_{i=1}^j
(d_i-1)$.
Collecting everything and the sum $\sum_{\sigma_j}$, on gets the
desired
result.

{\it b)} Two paths have a weight $a_j(j-1)!\sum_{k=2}^{j-2}{1\over
k}$.
First of all, there is a factor $(-2)^2$, due to the closed paths.
The first column gives $\left(\matrix{n-m\cr 2}\right)$, as before.
In the first row, I fix the second 1 in position
$(1,2)$, gaining a factor $j-1$. Now, consider the second column:
this time I
{\sl can} put a 1 in $(2,2)$,
as in the second case of (\ref{matri}).
This will close the first path, and only one path will remain to be
closed.
Now, consider the third column: with a factor $\left(\matrix{n-m-2\cr
2}\right)$,
I can always fix the two 1's in
$(3,3)$ and $(4,3)$ and proceed as in point {\it a)}. I shall have,
in total, a weight $a_j(j-1)(j-3)!=a_j{(j-1)!\over (j-2)}$.
If, instead, I do not close the first path by putting a 1 in position
$(2,2)$,
I can always arrange the second column by putting a 1 in $(3,2)$,
this giving
a factor $n-m-2$. Then,
I can close the first path on the third column,
for example, which is what happens in the third case of
(\ref{matri}).
This situation is characterized by
a weight $a_j{(j-1)!\over (j-3)}$, as it can be easily checked.
In other words I can close the first path in any column $k$ such that
$2\leq
k\leq j-2$.
This will produce a weight $a_j{(j-1)!\over (j-k)}$. Summing over the
various
possibilities, one gets the claimed weight for the two paths.

{\it c)} Reasoning in a completely similar way, one realizes that
three paths have a weight equal to
$-a_j(j-1)!\sum_{\sigma_2^{(j)}}\prod_{p\in\sigma_2^{(j)}}{1\over
p}$.
$\sigma_2^{(j)}$ stands for the couples $\{k,l\}$
such that $|k-l|>1$ (this is due to the fact that any closed path
occupies at least two columns) and $2\leq k,l\leq j-2$.
At this point, one easily learns the general rule and immediately
proves that
$k$ paths have a weight equal to
$(-1)^ka_j(j-1)!\sum_{\sigma_{k-1}^{(j)}}\prod_{p\in\sigma_{k-1}^{(j)}
}{1\over
p}$.

The total coefficient is thus simply $C^{(j)}_{\{d\}m}=c_ja_j$ with
\begin{equation}
c_j\equiv(j-1)!\sum_{k=1}^{\left[j/2\right]}
(-1)^k\sum_{\sigma_{k-1}^{(j)}}\prod_{p\in\sigma_{k-1}^{(j)}}{1\over
p}=
1-j.
\end{equation}
The last equality is easily proven by induction. Distinguishing those
$\sigma_{k-1}^{(j)}$'s
that contain $j-2$ from those that do not,
one finds the
following recursion relation:
\begin{equation}
c_j=(j-1)!\left(
\sum_{k=1}^{\left[{j-1\over 2}\right]}
(-1)^k\sum_{\sigma_{k-1}^{(j-1)}}\prod_{p\in\sigma_{k-1}^{(j-1)}}{1
\over p}+
{1\over j-2}\sum_{k=2}^{\left[{j\over 2}\right]}
(-1)^k\sum_{\sigma_{k-2}^{(j-2)}}\prod_{p\in\sigma_{k-2}^{(j-2)}}{1
\over
p}\right)=
(j-1)(c_{j-1}-c_{j-2}).
\end{equation}
The first values are
\begin{equation}
c_2=-1,\quad c_3=-2,\quad  c_4=3!\left(-1+{1\over 2}\right)=-3,\quad
c_5=4!\left(-1+{1\over 2}+{1\over 3}\right)=-4.
\end{equation}
This concludes the proof.

\end{document}